\documentclass[twocolumn,aps,pra,amsmath,amssymb]{revtex4-1}
\usepackage[latin9]{inputenc}
\usepackage{amsmath}
\usepackage{amssymb}
\usepackage{graphicx}
\usepackage{esint}

\makeatletter
\usepackage{epsfig}
\usepackage{bm}
\usepackage{dcolumn}
\usepackage{color}

\makeatother

\begin{document}
\title{Crossover polarons in a strongly interacting Fermi superfluid }
\author{Hui Hu$^{1}$, Jia Wang$^{1}$, Jing Zhou$^{2}$, and Xia-Ji Liu$^{1}$}
\affiliation{$^{1}$Centre for Quantum Technology Theory, Swinburne University
of Technology, Melbourne, Victoria 3122, Australia}
\affiliation{$^{2}$Department of Science, Chongqing University of Posts and Telecommunications,
Chongqing 400065, China}
\date{\today}
\begin{abstract}
We investigate the zero-temperature quasiparticle properties of a
mobile impurity immersed in a strongly interacting Fermi superfluid
at the crossover from a Bose-Einstein condensate (BEC) to a Bardeen--Cooper--Schrieffer
(BCS) superfluid, by using a many-body $T$-matrix approach that excludes
Efimov trimer bound states. Termed BEC-BCS crossover polaron, or crossover
polaron in short, this quasiparticle couples to elementary excitations
of a many-body background and therefore could provide a useful probe
of the underlying strongly interacting Fermi superfluid. Due to the
existence of a significant pairing gap $\Delta$, we find that the
repulsive polaron branch becomes less well-defined. In contrast, the
attractive polaron branch is protected by the pairing gap and becomes
more robust at finite momentum. It remains as a delta-function peak
in the impurity spectral function below a threshold $2\Delta$. Above
the threshold, the attractive polaron enters the particle-hole continuum
and starts to get damped. We predict the polaron energy, residue and
effective mass for realistic Bose-Fermi mixtures, where the minority
bosonic atoms play the role of impurity. These results are practically
useful for future cold-atom experiments on crossover polarons. 
\end{abstract}
\maketitle

\section{Introduction}

Polaron physics - quasiparticles formed by coupling a mobile impurity
to elementary excitations of a many-particle background - has received
increasing attention in the field of ultracold atoms over the last
fifteen years \citep{Massignan2014,Lan2014,Schmidt2018}. Owing to
the unprecedented controllability of interparticle interaction using
Feshbach resonances \citep{Bloch2008,Chin2010}, Fermi or Bose polaron,
namely, an impurity immersed in a non-interacting Fermi gas or a weakly
interacting Bose gas, has now been systematically explored in the
field of the cold-atom physics, both experimentally \citep{Schirotzek2009,Zhang2012,Kohstall2012,Koschorreck2012,Cetina2016,Hu2016,Jorgensen2016,Scazza2017,Zan2019,Zan2020,Ness2020}
and theoretically \citep{Chevy2006,Lobo2006,Combescot2007,Prokofev2008,Massignan2008,Combescot2008,Punk2009,Cui2010,Massignan2011,Mathy2011,Schmidt2012,Parish2013,Vlietinck2013,Rath2013,Doggen2013,Li2014,Kroiss2015,Levinsen2015,Hu2016PRA,Goulko2016,Hu2018,Tajima2018,Tajima2019,PenaArdila2019,Mulkerin2019,Liu2019,Wang2019,Parish2021,Isaule2021,Pessoa2021}.
A unique advantage of both Fermi and Bose polarons is their simplicity.
As the many-particle background is barely affected by the existence
of a single impurity, we may concentrate on the impurity only. As
a result, it is possible to make a quantitative comparison between
experimental data and theoretical predictions \citep{Massignan2014,Schmidt2018,Schirotzek2009,Scazza2017,Hu2018,Tajima2019,Parish2021,Isaule2021},
enabling us to examine in a stringent way different approximate quantum
many-particle theories \citep{Chevy2006,Combescot2007,Combescot2008,Liu2019}.

Another advantage of polarons is that they may provide a sensitive
probe of the background many-particle systems. In this respect, Fermi
or Bose polaron is not so useful, as the background non-interacting
Fermi gas or weakly interacting Bose gas is well understood. It would
be interesting to consider the polaron physics with a strongly correlated
background, such as a two-component Fermi superfluid near a Feshbach
resonance, which undergoes the crossover from a Bose-Einstein condensate
(BEC) to a Bardeen--Cooper--Schrieffer (BCS) superfluid \citep{Leggett1980,NSR1985,Hu2006}.
We dub the resulting quasiparticles as crossover polarons for convenience.
The experimental setup can be easily achieved by utilizing a superfluid
Bose-Fermi mixture, where the concentration of bosonic atoms can be
reduced to reach the polaron limit. Indeed, motivated by the recent
experimental demonstration of dual $^{6}$Li-$^{7}$Li \citep{FerrierBarbut2014},
$^{6}$Li-$^{41}$K \citep{Yao2016}, and $^{6}$Li-$^{174}$Yb superfluid
mixtures \citep{Roy2017}, a moving impurity immersed in a Fermi superfluid
has been considered by Nishida \citep{Nishida2015} and Yi and Cui
\citep{Yi2015} using a Chevy's variational ansatz, and more recently
by Pierce, Leyronas and Chevy \citep{Pierce2019} based on the second-order
perturbation theory. In these pioneering works, the role of three-body
Efimov physics has been highlighted and the unknown three-body parameter
is typically introduced through a large momentum cut-off \citep{Nishida2015,Yi2015}.

The modification of the polaron spectrum due to Efimov states is definitely
of great interest. However, realistically it is not clear whether
the trimer states can be tuned to be resonant with the polaron state
and whether these two kinds of states can have similar time-scale
in dynamics so both of them can be observed simultaneously. In Bose
polarons, the influence of Efimov states on the polaron physics seems
to be negligible for the typical gas parameter of the background BEC
(i.e., $na_{B}^{3}\sim10^{-5})$ \citep{Hu2016,Isaule2021}. It only
shows up for a highly compressible BEC when the gas parameter becomes
sufficiently small \citep{Jorgensen2016,Isaule2021}. As a Fermi superfluid
is less compressible than a typical BEC, naively one anticipates that
the interplay between Efimov and polaron physics might be difficult
to observe experimentally. In addition, the existence of Efimov trimers
in some realistic experiments might be unfavored: the impurity might
have a much larger mass or only strongly interact with one component
of the superfluid fermions. In order to understand the polaron physics
with a Fermi superfluid background in realistic experiments, it would
be useful to focus on the two-body sector and separate out the effects
of Efimov trimers.

The purpose of this work is to develop a theoretical framework for
the two-body sector of crossover polarons, based on a non-self-consistent
many-body $T$-matrix theory \citep{Combescot2007}. The use of the
$T$-matrix approach has two obvious advantages. On the one hand,
if we consider the BCS mean-field theory for the background Fermi
superfluid, the $T$-matrix approach has the same accuracy of Chevy's
variational ansatz adopted earlier \citep{Nishida2015,Yi2015}, but
enables us to concentrate on the two-body sector. On the other hand,
it allows to go beyond the qualitative BCS description of the strongly
interacting Fermi superfluid and therefore can predict quasiparticle
properties of crossover polarons in a quantitative manner.

In saying that, it is worth noting that a quantitative description
of the background many-body system of a strongly interacting Fermi
superfluid itself is a notoriously difficult problem \citep{Hu2006}.
Therefore, in this work, as a first attempt we would rather consider
the BCS description for the Fermi superfluid \citep{Leggett1980}
and focus on the non-trivial role played by the pairing gap on the
crossover polaron. We find that the repulsive polaron branch is less
favored by a significant pairing gap $\Delta$. For the attractive
polaron branch, the ground-state polaron energy at zero momentum increases
due to the existence of a threshold $2\Delta$ for dressing the impurity
with particle-hole excitations. However, the same threshold protects
the attractive polaron and makes it long-lived at larger momentum.
We predict the polaron properties for three Bose-Fermi mixtures in
the polaron limit, which might be readily examined in current cold-atom
laboratories.

\section{Many-body T-matrix approach}

We consider an impurity of mass $m_{I}$ moving in a bath of spin-1/2
Fermi superfluid of equal mass $m_{\uparrow}=m_{\downarrow}=m$, described
by the Hamiltonian $\mathcal{H}=\mathcal{H}_{I}+\mathcal{H}_{\textrm{int}}+\mathcal{H}_{\textrm{sf}}$.
Here, $\mathcal{H}_{I}$ and $\mathcal{H}_{\textrm{sf}}$ are the
Hamiltonian of the impurity and of the background Fermi superfluid,
respectively, and $\mathcal{H}_{\textrm{int}}$ describes the interaction
between the impurity and Fermi atoms. In the absence of the impurity-atom
interaction, the impurity has an energy spectrum of free particle,
i.e., $\epsilon_{\mathbf{p}}^{(I)}=\hbar^{2}\mathbf{p}^{2}/(2m_{I})$
at the momentum $\mathbf{p}$. In general, the model Hamiltonian of
a strongly interacting Fermi superfluid $\mathcal{H}_{\textrm{sf}}$
is difficult to solve. From now on, let us assume that it can be exactly
solved by the single-particle Green function $\mathcal{G}_{ij}(\mathcal{K}\equiv(\mathbf{k},i\omega_{m}))$
at the momentum $\mathbf{k}$ and fermionic Matsubara frequency $i\omega_{m}=(2m+1)\pi k_{B}T$,
which takes the form of a 2 by 2 matrix (i.e., $i,j=1,2=\uparrow,\downarrow$)
in accord with the use of the Nambu spinor $\Psi_{\mathbf{k}}=(\psi_{\mathbf{k}\uparrow},\psi_{-\mathbf{k}\downarrow}^{\dagger})^{T}$
for the atomic field operators in the broken-symmetry superfluid state.
The impurity-atom interaction can be conveniently described using
a contact potential (the system volume $V=1$),

\begin{equation}
\mathcal{H}_{\textrm{int}}=\sum_{\sigma=\uparrow,\downarrow}g_{\sigma}\sum_{\mathbf{q},\mathbf{k},\mathbf{k}'}\psi_{\mathbf{k}\sigma}^{\dagger}c_{\mathbf{q}-\mathbf{k}}^{\dagger}c_{\mathbf{q}-\mathbf{k}'}\psi_{\mathbf{k}'\sigma},\label{eq:intHami}
\end{equation}
where $c_{\mathbf{k}}$ and $c_{\mathbf{k}}^{\dagger}$ are respectively
annihilation and creation field operators for the impurity, and the
bare interaction strength $g_{\sigma}$ should be regularized by using
the $s$-wave scattering length $a_{\sigma}$ via the standard relation,
$g_{\sigma}^{-1}=m_{r}/(2\pi\hbar^{2}a_{\sigma})-\sum_{\mathbf{p}}2m_{r}/(\hbar^{2}\mathbf{p}^{2})$,
with the reduced mass $m_{r}\equiv mm_{I}/(m+m_{I})$.

In the non-self-consistent $T$-matrix approach for polarons \citep{Combescot2007},
we keep the ladder diagram for the successive particle-particle scattering
between the impurity and background fermionic atoms. However, the
use of the Nambu spinor mixes the particle-hole channel for atoms
\citep{Hu2006}. For example, the hole propagator of the spin-down
atoms actually represents the propagation of particle-like excitations.
This technical difficulty can be cured by taking the Green function
$-\mathcal{G}_{22}(-\mathcal{K})\equiv\mathcal{G}_{11}(\mathcal{K})$
as the particle-propagator for spin-down atoms. With this consideration,
we may directly write down the two-particle vertex functions $\Gamma_{ij}$
at the momentum $\mathbf{q}$ and Matsubara frequency $i\nu_{n}$
(collectively denoted as $\mathcal{Q}\equiv(\mathbf{q},i\nu_{n})$),
\begin{equation}
\Gamma\left(\mathcal{Q}\right)=\left[\begin{array}{cc}
\chi_{11}\left(\mathcal{Q}\right) & \chi_{12}\left(\mathcal{Q}\right)\\
\chi_{12}\left(\mathcal{Q}\right) & \chi_{22}\left(\mathcal{Q}\right)
\end{array}\right]^{-1},
\end{equation}
where the two-particle propagators $\chi_{ij}(\mathcal{Q})$ are given
by, 
\begin{eqnarray}
\chi_{11}\left(\mathcal{Q}\right) & = & \frac{1}{g_{\uparrow}}+\sum_{\mathcal{K}}\mathcal{G}_{11}\left(\mathcal{K}\right)G_{0}\left(\mathcal{Q}-\mathcal{K}\right),\label{eq:kappa11}\\
\chi_{12}\left(\mathcal{Q}\right) & = & \sum_{\mathcal{K}}\mathcal{G}_{12}\left(\mathcal{K}\right)G_{0}\left(\mathcal{Q}-\mathcal{K}\right),\label{eq:kappa12}\\
\chi_{22}\left(\mathcal{Q}\right) & = & \frac{1}{g_{\downarrow}}+\sum_{\mathcal{K}}\left[-\mathcal{G}_{22}\left(-\mathcal{K}\right)\right]G_{0}\left(\mathcal{Q}-\mathcal{K}\right),\label{eq:kappa22}
\end{eqnarray}
$\sum_{\mathcal{K}}$ or $\sum_{\mathcal{Q}}$ stands for the short-hand
notation $k_{B}T\sum_{i\omega_{m}}\sum_{\mathbf{k}}$ or $k_{B}T\sum_{i\nu_{n}}\sum_{\mathbf{q}}$,
and $G_{0}(\mathcal{Q}-\mathcal{K})=1/[i\nu_{n}-i\omega_{m}-\epsilon_{\mathbf{q}-\mathbf{k}}^{(I)}]$
is the non-interacting Green function of the impurity. The self-energy
$\Sigma(\mathcal{K})=\Sigma_{11}+2\Sigma_{12}+\Sigma_{22}$ of the
impurity then takes the form \citep{Combescot2007}, 
\begin{eqnarray}
\Sigma_{11}\left(\mathcal{K}\right) & = & +\sum_{\mathcal{Q}}\Gamma_{11}\left(\mathcal{Q}\right)\mathcal{G}_{11}\left(\mathcal{Q}-\mathcal{K}\right),\\
\Sigma_{12}\left(\mathcal{K}\right) & = & -\sum_{\mathcal{Q}}\Gamma_{12}\left(\mathcal{Q}\right)\mathcal{G}_{12}\left(\mathcal{Q}-\mathcal{K}\right),\\
\Sigma_{22}\left(\mathcal{K}\right) & = & +\sum_{\mathcal{Q}}\Gamma_{22}\left(\mathcal{Q}\right)\left[-\mathcal{G}_{22}\left(\mathcal{-Q}+\mathcal{K}\right)\right],
\end{eqnarray}
where the different sign in $\Sigma_{11}$ (or $\Sigma_{22}$) and
$\Sigma_{12}$ is due to the absence of a Fermi loop in the diagram
for $\Sigma_{12}$.

\begin{figure*}
\begin{centering}
\includegraphics[width=0.9\textwidth]{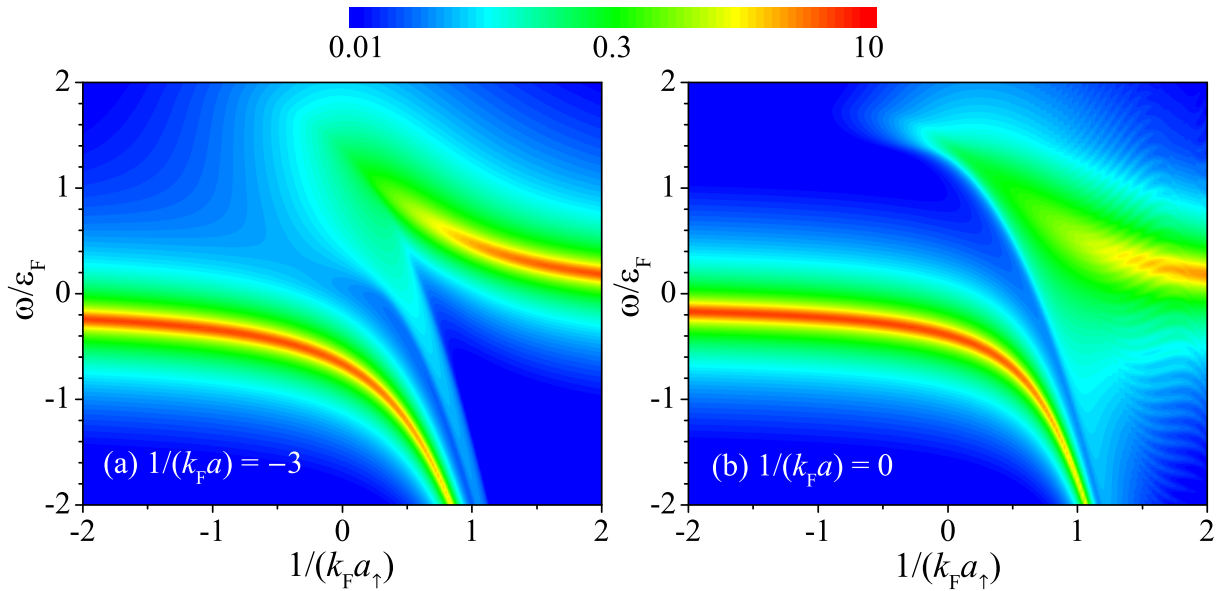}
\par\end{centering}
\caption{\label{fig1_akwaif} The evolution of the zero-momentum impurity spectral
function $A(\mathbf{k}=0,\omega$) as a function of the scattering
length $a_{\uparrow}$ between the impurity and spin-up atoms, with
a weakly-interacting Fermi gas background (i.e., $1/(k_{F}a)=-3$
in (a)) or with a unitary Fermi superfluid background ($1/(k_{F}a)=0$
in (b)). The interaction parameter between the impurity and spin-down
atoms is set to be $1/(k_{F}a_{\downarrow})=-5$. A small linewidth
broadening factor $\eta=0.05\varepsilon_{F}$ has been introduced
to better illustrate the $\delta$-function peak of attractive polaron
branch. The spectral function is in units of $\varepsilon_{F}^{-1}$,
where $\varepsilon_{F}\equiv\hbar^{2}k_{F}^{2}/(2m)$ is the Fermi
energy. We always take the equal mass for the impurity and atoms,
$m_{I}=m$, except the realistic experimental case in Fig. \ref{fig6_experiment}.}
\end{figure*}

\subsection{BCS Fermi superfluid}

The above expressions for the two-particle vertex function and impurity
self-energy are quantitatively useful, provided that the Green functions
$\mathcal{G}_{ij}(\mathcal{K})$ of the background Fermi superfluid
are known to a certain accuracy. Here, we are interested in understanding
the general picture of the crossover polaron, with the help of the
qualitatively reliable mean-field theory for Fermi superfluids. The
consideration of strong pair fluctuations to $\mathcal{G}_{ij}(\mathcal{K})$
at the BEC-BCS crossover is postponed to a future study.

In the mean-field framework, the single-particle Green functions $\mathcal{G}_{ij}(\mathcal{K})$
are well-known \citep{Leggett1980,Hu2006}: 
\begin{eqnarray}
\mathcal{G}_{11}\left(\mathcal{K}\right) & = & \frac{u_{\mathbf{k}}^{2}}{i\omega_{m}-E_{\mathbf{k}}}+\frac{v_{\mathbf{k}}^{2}}{i\omega_{m}+E_{\mathbf{k}}},\\
\mathcal{G}_{12}\left(\mathcal{K}\right) & = & \frac{u_{\mathbf{k}}v_{\mathbf{k}}}{i\omega_{m}-E_{\mathbf{k}}}+\frac{u_{\mathbf{k}}v_{\mathbf{k}}}{i\omega_{m}+E_{\mathbf{k}}},
\end{eqnarray}
where $E_{\mathbf{k}}\equiv\sqrt{\xi_{\mathbf{k}}^{2}+\Delta^{2}}$
with $\xi_{\mathbf{k}}\equiv\epsilon_{\mathbf{k}}-\mu=\hbar^{2}\mathbf{k}^{2}/(2m)-\mu$
is the Bogoliubov quasiparticle energy for a Fermi superfluid with
chemical potential $\mu$ and pairing gap $\Delta$, and $u_{\mathbf{k}}^{2}=[1+\xi_{\mathbf{k}}/E_{\mathbf{k}}]/2$,
$v_{\mathbf{k}}^{2}=1-u_{\mathbf{k}}^{2}$, and $u_{\mathbf{k}}v_{\mathbf{k}}=\Delta/(2E_{\mathbf{k}})$
are the quasiparticle wavefunctions. Both $\mu$ and $\Delta$ can
be calculated for a given dimensionless interaction parameter $1/(k_{F}a)$
\citep{Leggett1980}, where $a$ is the $s$-wave length between unlike
atoms and $k_{F}=(3\pi^{2}n)^{1/3}$ is Fermi wavevector at the number
density $n$ of the Fermi superfluid. After plugging the above $\mathcal{G}_{ij}(\mathcal{K})$
and the free impurity Green function $G_{0}(\mathcal{Q}-\mathcal{K})$
into the expressions of the two-particle propagators, we find that
at zero temperature, 
\begin{eqnarray}
\chi_{dd}\left(\mathcal{Q}\right) & = & -\sum_{\mathbf{p}}\left[\frac{u_{\mathbf{p}}^{2}}{i\nu_{n}-E_{\mathbf{p}}-\epsilon_{\mathbf{q}-\mathbf{p}}^{(I)}}+\frac{2m_{r}}{\hbar^{2}\mathbf{p}^{2}}\right],\\
\chi_{12}\left(\mathcal{Q}\right) & = & -\sum_{\mathbf{p}}\frac{u_{\mathbf{p}}v_{\mathbf{p}}}{i\nu_{n}-E_{\mathbf{p}}-\epsilon_{\mathbf{q}-\mathbf{p}}^{(I)}},
\end{eqnarray}
and $\chi_{11}(\mathcal{Q})=m_{r}/(2\pi\hbar^{2}a_{\uparrow})+\chi_{dd}(\mathcal{Q})$
and $\chi_{22}(\mathcal{Q})=m_{r}/(2\pi\hbar^{2}a_{\downarrow})+\chi_{dd}(\mathcal{Q})$.
Using the fact that there is no macroscopic population of the impurity
state, we can analytically perform the Matsubara frequency summation
in the expressions of $\Sigma_{ij}(\mathcal{K})$ \citep{Combescot2007}.
Thus, after analytic continuation ($i\omega_{m}\rightarrow\omega^{+}\equiv\omega+i0^{+}$)
we obtain the zero-temperature retarded self-energies, 
\begin{eqnarray}
\Sigma_{11}\left(\mathbf{k},\omega\right) & = & +\sum_{\mathbf{q}}v_{\mathbf{q}-\mathbf{k}}^{2}\Gamma_{11}\left(\mathbf{q},\omega^{+}-E_{\mathbf{q}-\mathbf{k}}\right),\\
\Sigma_{12}\left(\mathbf{k},\omega\right) & = & -\sum_{\mathbf{q}}u_{\mathbf{q}-\mathbf{k}}v_{\mathbf{q}-\mathbf{k}}\Gamma_{12}\left(\mathbf{q},\omega^{+}-E_{\mathbf{q}-\mathbf{k}}\right),\\
\Sigma_{22}\left(\mathbf{k},\omega\right) & = & +\sum_{\mathbf{q}}v_{\mathbf{q}-\mathbf{k}}^{2}\Gamma_{22}\left(\mathbf{q},\omega^{+}-E_{\mathbf{q}-\mathbf{k}}\right).
\end{eqnarray}
The retarded interacting impurity Green function is then given by
$G(\mathbf{k},\omega)=1/[\omega-\epsilon_{\mathbf{k}}^{(I)}-\Sigma(\mathbf{k},\omega)]$,
and its pole position determines the polaron energy, i.e., 
\begin{equation}
\mathcal{E}_{P}\left(\mathbf{k}\right)=\epsilon_{\mathbf{k}}^{(I)}+\Sigma\left[\mathbf{k},\mathcal{E}_{P}\left(\mathbf{k}\right)\right].\label{eq:PolaronEnergy}
\end{equation}
By expanding the self-energy near the zero momentum $\mathbf{k}=0$
and the ground-state polaron energy $\mathcal{E}_{P}\equiv\mathcal{E}_{P}\left(\mathbf{0}\right)$,
we can calculate directly the polaron residue $\mathcal{Z}^{-1}=1-\partial\textrm{Re}\Sigma(0,\omega)/\partial\omega$
and the effective mass $m_{*}/m_{I}=\mathcal{Z}^{-1}/[1+\partial\textrm{Re}\Sigma(\mathbf{k},\mathcal{E}_{P})/\partial\epsilon_{\mathbf{k}}^{(I)}]$.

\subsection{Link to Chevy's ansatz}

At this stage, it is beneficial to clarify the connection of our many-body
$T$-matrix formalism to Chevy's variational ansatz \citep{Chevy2006,Combescot2007}.
Let us focus on the simplest case with zero interaction between the
impurity and spin-down atoms ($a_{\downarrow}=0^{-}$), where $\chi_{22}\rightarrow-\infty$
so the only remaining two-particle vertex function is $\Gamma_{11}$.
It is readily seen that the equation for the polaron energy Eq. (\ref{eq:PolaronEnergy})
can be rewritten as, 
\begin{eqnarray}
\mathcal{E}_{P}\left(\mathbf{k}\right) & = & \epsilon_{\mathbf{k}}^{(I)}+\sum_{\mathbf{q}}v_{\mathbf{q}}^{2}\left[\frac{m_{r}}{2\pi\hbar^{2}a_{\uparrow}}-\sum_{\mathbf{p}}\frac{2m_{r}}{\hbar^{2}\mathbf{p}^{2}}\right.\nonumber \\
 &  & \left.+\sum_{\mathbf{p}}\frac{u_{\mathbf{p}}^{2}}{E_{\mathbf{p}}+\epsilon_{\mathbf{k}+\mathbf{q}-\mathbf{p}}^{(I)}+E_{\mathbf{q}}-\mathcal{E}_{P}\left(\mathbf{k}\right)}\right]^{-1}.\label{eq:PolaronEnergy2}
\end{eqnarray}
In the case of an ideal Fermi gas background with a vanishing pairing
gap $\Delta=0$, we have $v_{\mathbf{q}}^{2}=\Theta(k_{F}-q)$, $u_{\mathbf{p}}^{2}=\Theta(p-k_{F})$,
$E_{\mathbf{p}}=\epsilon_{\mathbf{p}}-\mu$ and $E_{\mathbf{q}}=\mu-\epsilon_{\mathbf{q}}$,
where $\Theta(x)$ is the Heaviside step function. Thus, we recover
the celebrated equation for polaron energy from Chevy's variational
ansatz (see Eq. (2) in the seminal work \citep{Combescot2007}). For
the alternative derivation of Eq. (\ref{eq:PolaronEnergy2}) by using
Chevy's ansatz with the standard BCS variational wave-function, we
refer to Appendix A.

\subsection{Remarks on the BEC limit}

The above formalisms are quantitatively applicable to the BCS limit
and are qualitatively applicable in the unitary or crossover regime.
Towards the BEC limit, the fermionic degree of freedom, as characterized
by the Green functions $\mathcal{G}_{ij}(\mathcal{K})$, is suppressed.
The new bosonic degree of freedom, represented by the vertex function
of the Fermi superfluid \cite{NSR1985,Hu2006}, become dominant. As
a result, our formalisms based on the non-self-consistent $T$-matrix
theory fail. We need to construct a theory beyond the ladder approximation,
by considering the exact \emph{three-body} interaction vertex involving
a Cooper pair and the impurity that describes the dressing of the
impurity with the \emph{gapless} phonon excitations of the Fermi superfluid
\cite{Hu2006}. In this way, we are anticipated to recover Bose polarons
occurring in the weakly-interacting BEC limit \cite{Rath2013,PenaArdila2019}.

\begin{figure}[t]
\begin{centering}
\includegraphics[width=0.48\textwidth]{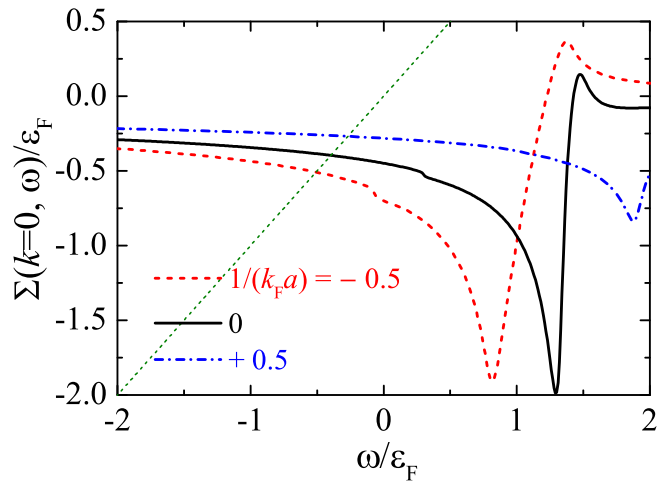}
\par\end{centering}
\caption{\label{fig2_selfenergy} The zero-momentum impurity self-energy at
different background Fermi superfluids characterized by the interaction
parameter, $1/(k_{F}a)=-0.5$ (red dashed line), $0$ (black solid
line), and $+0.5$ (blue dot-dashed line). The thin green dashed line
shows the straight line $h(\omega)=\omega$. The intersection of the
straight line and the self-energy curve determines the polaron energy
at zero momentum. Here, we take the impurity-atom interactions: $1/(k_{F}a_{\uparrow})=0$
and $a_{\downarrow}=0^{-}$. The two times pairing gap for each curve
is: $2\Delta\simeq0.804\varepsilon_{F}$ at $1/(k_{F}a)=-0.5$, $2\Delta\simeq1.373\varepsilon_{F}$
in the unitary limit, and $2\Delta\simeq2.033\varepsilon_{F}$ at
$1/(k_{F}a)=+0.5$.}
\end{figure}

\section{Results and discussions}

Before we present the results, let us briefly discuss how the paired
Fermi superfluid background is affected by the moving impurity. Dynamically,
if we take a snapshot, an impurity will excite density fluctuations
of the total density (i.e., charge degree of freedom) and of the spin
density (i.e., spin degree of freedom) at the impurity site, which
propagate over the whole Fermi superfluid. As the impurity moves,
the fluctuations generated at different time interfere and after a
timescale set by the inverse Fermi energy the fluctuations fade away.
In the limit of a single impurity, therefore in equilibrium the Fermi
superfluid is essentially not perturbed, since the perturbation strength
scales like $1/N$, where $N$ is the total number of fermions of
the background system.

The situation may dramatically change if the mass of the impurity
is very large. For example, an infinitely heavy impurity will create
a \emph{static} scattering potential and therefore lead to a permanent
local distortion of the Fermi superfluid near the impurity. For a
static non-magnetic impurity scattering (i.e., $a_{\uparrow}=a_{\downarrow}$),
the single-particle energy spectrum of the Fermi superfluid is essentially
unchanged, according to Anderson's theorem \cite{Anderson1959,Balatsky2006}.
For a static magnetic impurity scattering ($a_{\uparrow}\neq a_{\downarrow}$)
that lifts the Kramers degeneracy of the pairing states (or breaks
the time-reversal symmetry) and excites spin-density fluctuations,
a non-trivial Yu-Shiba-Rusinov (YSR) bound state appears within the
energy gap $2\Delta$ \cite{Yu1965,Shiba1968,Rusinov1969,Vernier2011,Jiang2011}. 

The existence of the YSR bound state in the case of a moving impurity
requires a further investigation. Naïvely, we may anticipate the appearance
of a sub-gap band of YSR states, whose number is proportional to the
number of impurities or impurity density (which scales to zero in
the single impurity limit). In our non-self-consistent $T$-matrix
approach, we consider the single-impurity limit and hence neglect
again the coupling of the YSR bound state to the polaron in the \emph{thermodynamic}
limit. Nevertheless, in the static limit the impurity properties could
be profoundly affected by the YSR bound state, since we probe exactly
the neighborhood of the impurity (instead of the whole system for
a moving impurity). 

\subsection{The significance of a pairing gap}

Appendix A shows that Eq. (\ref{eq:PolaronEnergy2}) can be understood
as an extension of Chevy's ansatz in the case of a Fermi superfluid
background. It describes the virtual process of dressing the impurity
with simultaneous particle- and hole-like excitations, each of which
has the possibility $u_{\mathbf{p}}^{2}$ or $v_{\mathbf{q}}^{2}$,
and energy $E_{\mathbf{p}}$ or $E_{\mathbf{q}}$. The appearance
of the summation $E_{\mathbf{p}}+E_{\mathbf{q}}$ in the denominator
of Eq. (\ref{eq:PolaronEnergy2}) is a direct consequence of breaking
a Cooper pair during the virtual excitation. It naturally leads to
a threshold $2\Delta$ for the one-particle-hole excitation, which
may change the polaron spectrum in a non-trivial way.

To see this, in Fig. 1 we show the impurity spectral function $A(\mathbf{k},\omega)\equiv-\textrm{(1/\ensuremath{\pi})Im}G(\mathbf{k},\omega)$
at zero momentum $\mathbf{k}=0$ as a function of the dimensionless
impurity-atom interaction $1/(k_{F}a_{\uparrow})$. For a negligible
pairing gap in (a), we find the typical spectrum for a Fermi polaron
with both attractive branch and repulsive branch \citep{Massignan2011},
and a narrow molecule-hole continuum in between \citep{Massignan2011,Parish2021}.
There are notable changes in the spectrum when we take a unitary Fermi
gas as the background with a significant (mean-field) pairing gap
$\Delta\simeq0.69\varepsilon_{F}$. The repulsive branch becomes much
blurred, indicating the shorter lifetime of repulsive polarons. At
the same time, the parameter window for their existence shrinks. The
molecule-hole continuum also disappears. In contrast, the attractive
branch remains well-defined, but the ground-state energy of attractive
polarons seems to increase systematically.

As the repulsive polaron can be naïvely viewed as the excited state
of the molecule (with the dressing of particle-hole excitations),
the less well-defined repulsive polaron may be understood from the
fact that the molecule consisting of the impurity and spin-up atoms
becomes more difficult to form due to the energy cost for Cooper pair-breaking.
This idea is consistent with the disappearance of the molecule-hole
continuum. 

On the other hand, the upshift of the attractive polaron energy can
be easily understood from the particle-hole excitation threshold $2\Delta$,
which leads to a larger self-energy for the impurity, as shown in
Fig. \ref{fig2_selfenergy}, where we plot the zero-momentum self-energy
at various superfluid pairing gaps and at the unitary coupling between
the impurity and spin-up atoms (i.e., $1/(k_{F}a_{\uparrow})=0$).
By increasing the background interaction parameter $1/(k_{F}a)$ and
hence the pairing gap, the impurity self-energy shifts up at the negative
frequency. As a result, the polaron energy of the solution $\mathcal{E}_{P}=\Sigma(0,\mathcal{E}_{P})$,
which is given by the intersection of the straight line $h(\omega)=\omega$
and the self-energy curve $\Sigma(0,\omega)$, becomes larger.

\begin{figure}[t]
\begin{centering}
\includegraphics[clip,width=0.48\textwidth]{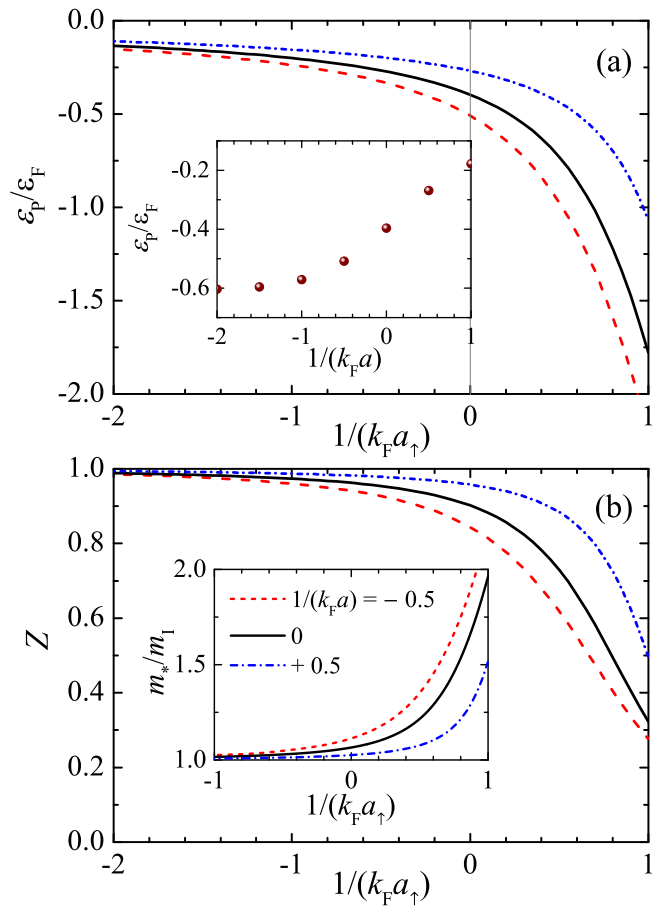} 
\par\end{centering}
\caption{\label{fig3_polaron} Polaron energy (a), residue (b) and effective
mass (the inset in (b)) as a function of the dimensionless impurity-atom
interaction parameter $1/(k_{F}a_{\uparrow})$ at three different
background Fermi superfluids with $1/(k_{F}a)=-0.5$ (red dashed line),
$0$ (black solid line), and $+0.5$ (blue dot-dashed line). The inset
in (a) shows the polaron energy as a function of $1/(k_{F}a)$ at
the unitary coupling $1/(k_{F}a_{\uparrow})=0$. For all the plots
in the figure, we consider the non-interacting limit between the impurity
and spin-down atoms (i.e., $a_{\downarrow}=0^{-}$).}
\end{figure}

\subsection{Equal mass case}

From now on, let us concentrate on well-defined attractive polarons,
which are easier to measure in experiments \citep{Schirotzek2009}.
In Fig. \ref{fig3_polaron}(a), we present the polaron energy as a
function of $1/(k_{F}a_{\uparrow})$ at $a_{\downarrow}=0^{-}$ and
at three different Fermi superfluid background. The increase of the
polaron energy as a function of the background interaction parameter
$1/(k_{F}a)$ is highlighted in the inset, where we take a unitary
coupling ($a_{\uparrow}\rightarrow\pm\infty$) between the impurity
and spin-up atoms. This is the most interesting case, in which we
may define a universal energy parameter \citep{Hu2006} for crossover
polarons, i.e., $\mathcal{E}_{P}=\xi(1/k_{F}a)\varepsilon_{F}$. From
the inset, we find that towards the BCS limit of Fermi superfluid
the energy parameter $\xi$ quickly saturates to the well-known result
for Fermi polarons, i.e., $\xi(a=0^{-})\simeq-0.607$ \citep{Chevy2006,Combescot2007}.
For a unitary Fermi superfluid background, it increases to $\xi(a=\pm\infty)\simeq-0.396$.
Although the difference between these two values $\Delta\xi=\xi(a=\pm\infty)-\xi(a=0^{-})\simeq0.211$
is not significant, it is slightly larger than the experimental resolution
in determining the polaron energy via the radio-frequency spectroscopy
(i.e., $\sim0.1\varepsilon_{F}$) \cite{Schirotzek2009,Scazza2017,Zan2019}
and therefore might be experimentally measured. We note that, the
difference $\Delta\xi$ may also quantitatively change, if we go beyond
the mean-field treatment of the strongly interacting Fermi superfluid
background. 

In line with the decreasing absolute value of the polaron energy within
a strongly interacting Fermi superfluid, the residue and effective
mass of the polaron increases and decreases, respectively, as reported
in Fig. \ref{fig3_polaron}(b). In particular, at the unitary impurity-atom
coupling ($a_{\uparrow}=\pm\infty$), we obtain a residue $Z\simeq0.90$
and an effective mass $m_{*}/m_{I}\simeq1.07$ with a unitary Fermi
superfluid, in comparison to the predictions of $Z\simeq0.78$ and
$m_{*}/m_{I}\simeq1.17$ in the case of Fermi polarons \citep{Combescot2007}.

\begin{figure}[t]
\begin{centering}
\includegraphics[width=0.48\textwidth]{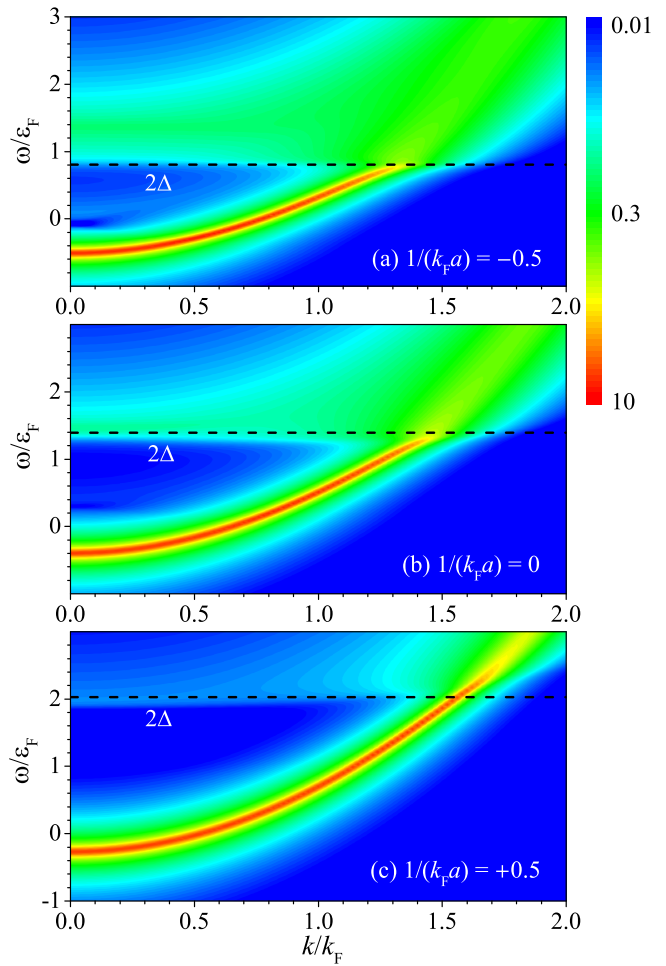} 
\par\end{centering}
\caption{\label{fig4_akw2d} Two-dimensional contour plot of the impurity spectral
function $A(k,\omega)$ at three different background Fermi superfluids
with $1/(k_{F}a)=-0.5$ (a), $0$ (b), and $+0.5$ (c). We consider
the unitary coupling between the impurity and spin-up atoms ($a_{\uparrow}=\pm\infty$)
and zero coupling between the impurity and spin-down atoms ($a_{\downarrow}=0^{-}$).
The black dashed lines indicate the particle-hole excitation thresholds,
$2\Delta\simeq0.804\varepsilon_{F}$ in (a) $2\Delta\simeq1.373\varepsilon_{F}$
in (b) and $2\Delta\simeq2.033\varepsilon_{F}$ in (c). We use a small
linewidth broadening factor $\eta=0.05\varepsilon_{F}$ to better
show the $\delta$-peak of attractive polaron branch. The spectral
function is in units of $\varepsilon_{F}^{-1}$.}
\end{figure}

We consider so far the ground-state polaron at zero momentum. At finite
momentum, in general the polaron will have a finite lifetime, once
its energy reaches the minimum energy of the particle-hole continuum.
For Fermi polarons, this occurs at zero energy $\omega=0$ in the
absence of two-body molecular bound states (i.e., $a_{\uparrow}\leq0$).
For crossover polarons with a significant background pairing gap $\Delta\sim O(\varepsilon_{F})$,
Eq. (\ref{eq:PolaronEnergy2}) indicates that we have a threshold
$2\Delta$ for particle-hole excitations and therefore the crossover
polaron should remain as a long-lived quasiparticle as long as its
energy is smaller than $2\Delta$. This anticipation is examined in
Fig. \ref{fig4_akw2d}, where we present two-dimensional contour plots
of the finite-momentum impurity spectral function $A(\mathbf{k},\omega)$
at three background Fermi superfluids. The threshold $2\Delta$ has
been shown in the figures by using black dashed lines. It is readily
seen that the polaron remains as a $\delta$-function peak, provided
that $\mathcal{E}_{P}+\hbar^{2}k^{2}/(2m_{*})<2\Delta$. This gives
rise to a characteristic momentum $k_{c}=\sqrt{2m^{*}(2\Delta-\mathcal{E}_{P})}/\hbar$,
below which the polaron is long-lived. We find that $k_{c}$ increases
with increasing pairing gap, suggesting the polaron becomes more robust
with a strongly interacting Fermi superfluid background. Finally,
if the impurity spectral function can be experimentally measured by
the \emph{momentum-resolved} radio-frequency spectroscopy, one may
directly determine the threshold $2\Delta$. This provides a possible
way to measure the pairing gap of the Fermi superfluid background.

\begin{figure}[t]
\begin{centering}
\includegraphics[width=0.48\textwidth]{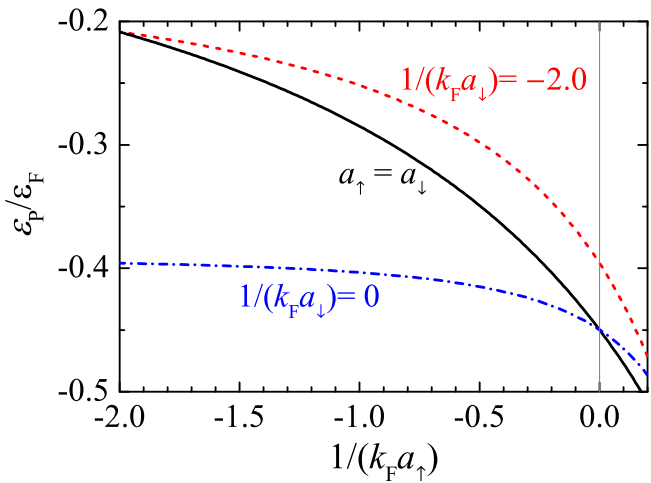} 
\par\end{centering}
\caption{\label{fig5_polaronvxup} The effect of the interaction between the
impurity and spin-down atoms on the polaron energy. Here, we consider
a unitary Fermi superfluid as the background ($a=\pm\infty$) and
three cases for the interaction between the impurity and spin-down
atom: $1/(k_{F}a_{\downarrow})=-0.5$ (red dashed line), $1/(k_{F}a_{\uparrow})$
(black solid line), and $0$ (blue dot-dashed line).}
\end{figure}

\subsection{Role of the interaction with spin-down atoms}

In previous calculations, we take a negligible interaction between
the impurity and spin-down atoms. This is the typical situation in
experiments since it is difficult to tune the two scattering lengths
$a_{\uparrow}$ and $a_{\downarrow}$ to be large simultaneously.
In Fig. \ref{fig5_polaronvxup}, we report the exceptional case that
the impurity interacts strongly with both spins. At the unitary coupling
($a_{\uparrow}=a_{\downarrow}=\pm\infty$) with a unitary Fermi superfluid
background ($a=\pm\infty$), we find the polaron energy $\mathcal{E}_{P}\simeq-0.450\varepsilon_{F}$.

\begin{figure}
\begin{centering}
\includegraphics[width=0.48\textwidth]{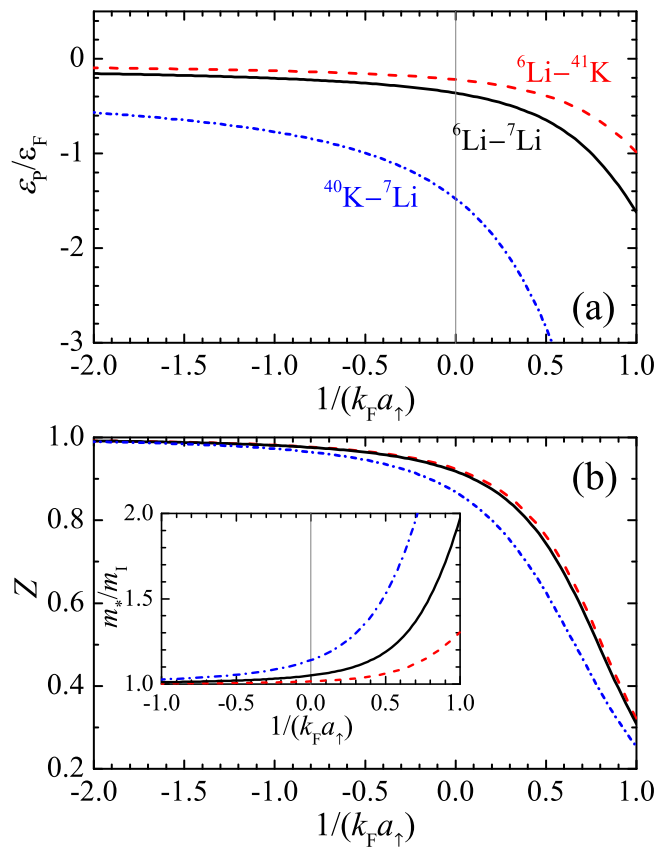} 
\par\end{centering}
\caption{\label{fig6_experiment} Polaron energy (a), residue (b) and effective
mass (the inset in (b)) as a function of the dimensionless impurity-atom
interaction parameter $1/(k_{F}a_{\uparrow})$ for the mixtures of
$^{6}$Li-$^{41}$K (red dashed line), $^{6}$Li-$^{7}$Li (black
line) and $^{40}$K-$^{7}$Li (blue dot-dashed line). We consider
a unitary Fermi superfluid as the background ($a=\pm\infty$) and
take $1/(k_{F}a_{\downarrow})=-5$ for the interaction between the
impurity and spin-down atoms.}
\end{figure}

\subsection{Experimental relevance}

Let us finally explore the possibility of experimentally observing
the predicted crossover polaron. A straightforward idea is to use
the recently realized dual Bose-Fermi superfluid \citep{FerrierBarbut2014,Yao2016,Roy2017}.
In the limit of small bosonic density, the bosons can be treated as
independent impurities and their properties can be directly measured
by using rf \citep{Schirotzek2009} or Raman spectroscopy \citep{Ness2020}.
Fermi-Fermi mixtures involving a (majority) two-component Fermi superfluid
and another (minority) normal Fermi gas, such as a paired $^{6}$Li
superfluid with $^{40}$K atoms as impurities, may also be possible
candidates. 

In Fig. \ref{fig6_experiment}, we show the polaron energy, residue
and effective mass in three Bose-Fermi mixtures with different mass
ratio $m_{I}/m$. For a lighter impurity, quasiparticle properties
appear to have a more sensitive dependence on the impurity-atom interaction. 

The experimental challenge of realizing a crossover polaron lies in
the difficulty of independently tuning the impurity-atom interaction
$a_{\uparrow}$, in addition to the control of the scattering length
$a$ for the background Fermi superfluid. This would require a significant
overlap between two Feshbach resonances for enlarging $a_{\uparrow}$
and $a$, respectively. A careful search for the best candidate system,
through detailed two-body calculations for various $s$-wave scattering
lengths, will be addressed in future works. 

\section{Conclusions and outlooks}

In summary, we have presented a many-body $T$-matrix theory for a
novel type of crossover polaron, which can be potentially realized
in dual Bose-Fermi superfluids, where the minority bosonic atoms can
be treated as impurities. By using a mean-field description for the
background strongly interacting Fermi superfluid, we have qualitatively
clarified the role played by the pairing gap on polaron physics. We
have found that the repulsive polaron branch ceases to exist. In contrast,
attractive polarons become robust at finite momentum. In the near
future, we will investigate quantitative corrects to the polaron quasiparticle
properties due to strong pair fluctuations in the background Fermi
superfluid. Another interesting possibility is to consider a topological
Fermi superfluid as the background and to study how the polaron physics
changes near the topological phase transition \citep{Bai2018}.

In our work, we do not consider the Efimov trimer bound states, which
may become important under specific conditions (with fine-tuning interaction
parameters). To address these trimer states and their interplay with
polaron states, we need to go beyond the current non-self-consistent
many-body $T$-matrix approach as discussed in Sec. IIC. A simple
way is to include the non-trivial vertex correction to the vertex
function $\Gamma(\mathcal{Q})$. In other words, we need to consider
the contribution of particle-hole diagrams to $\Gamma(\mathcal{Q})$,
together with the particle-particle ladder diagrams as given in Eqs.
(\ref{eq:kappa11}), (\ref{eq:kappa12}) and (\ref{eq:kappa22}).
The particle-hole diagrams take into account the multiple particle-hole
excitations near Fermi surfaces and capture the contribution of Efimov
states \cite{Punk2009,Nishida2015,Yi2015,Pierce2019}. This non-trivial
extension of the many-body $T$-matrix theory will be considered in
future studies.
\begin{acknowledgments}
We are grateful to Xing-Can Yao for useful discussions. This research
was supported by the Australian Research Council's (ARC) Discovery
Program, Grants No. DE180100592 and No. DP190100815 (J.W.), and Grant
No. DP180102018 (X.-J.L).
\end{acknowledgments}

\appendix
%dummy comment inserted by tex2lyx to ensure that this paragraph is not empty%dummy comment inserted by tex2lyx to ensure that this paragraph is not empty

\section{Variational Ansatz Approach}

Here we give the details on deriving the polaron energy relation,
i.e., Eq. (17) in the main text, by using an extended Chevy's ansatz
approach with the standard BCS variational wave-functions for the
case of a vanishing interaction $g_{\downarrow}=0$ between the impurity
and spin-down fermions.

The Hamiltonian is given by $\mathcal{H}=\mathcal{H}_{I}+\mathcal{H}_{{\rm int}}+\mathcal{H}_{{\rm sf}}$.
The first term is the Hamiltonian for the impurity, $\mathcal{H}_{I}=\sum_{\mathbf{k}}\epsilon_{\mathbf{k}}^{(I)}c_{\mathbf{k}}^{\dagger}c_{\mathbf{k}}$,
with $\epsilon_{\mathbf{k}}^{(I)}=\hbar^{2}\mathbf{k}^{2}/(2m_{I})$.
Here, $c_{\mathbf{k}}^{\dagger}$ and $c_{\mathbf{k}}$ are the annihilation
and creation field operators for the impurity, respectively. The impurity-fermion
interaction is given by (the system volume $V=1$) 
\begin{equation}
\mathcal{H}_{\text{int }}=g_{\uparrow}\sum_{\mathbf{q},\mathbf{k},\mathbf{k}^{\prime}}\psi_{\mathbf{k}\uparrow}^{\dagger}c_{\mathbf{q}-\mathbf{k}}^{\dagger}c_{\mathbf{q}-\mathbf{k}^{\prime}}\psi_{\mathbf{k}^{\prime}\uparrow}
\end{equation}
for the $g_{\downarrow}=0$ case, and the Hamiltonian for the two-component
Fermi superfluid is given by
\begin{widetext}
\begin{equation}
\mathcal{H}_{{\rm sf}}=\sum_{\sigma=\uparrow,\downarrow}\sum_{\mathbf{k}}\left(\epsilon_{\mathbf{k}}-\mu\right)\psi_{\mathbf{k}\sigma}^{\dagger}\psi_{\mathbf{k}\sigma}+g\sum_{\mathbf{k},\mathbf{k}^{\prime},\mathbf{q}}\psi_{\mathbf{k}\uparrow}^{\dagger}\psi_{\mathbf{q}-\mathbf{k}\downarrow}^{\dagger}\psi_{\mathbf{q}-\mathbf{k}^{\prime}\downarrow}\psi_{\mathbf{k}^{\prime}\uparrow},
\end{equation}
where $\epsilon_{\mathbf{k}}=\hbar^{2}\mathbf{k}^{2}/(2m)$ is the
dispersion relation for a free fermion; $\mu$ is the chemical potential;
$\psi_{\sigma}^{\dagger}$ and $\psi_{\sigma}$ are the annihilation
and creation field operators for the $\sigma$-component ($\sigma=\uparrow,\downarrow$)
fermion, respectively. The bare interaction strength $g_{\uparrow}$
(or $g$) should be regularized by using the $s$-wave scattering
length $a_{\uparrow}$ (or $a$) via the standard regularization relation.
For example, $g_{\uparrow}^{-1}=m_{r}/\left(2\pi\hbar^{2}a_{\uparrow}\right)-\sum_{\mathbf{p}}2m_{r}/\left(\hbar^{2}\mathbf{p}^{2}\right)$,
with the reduced mass $m_{r}\equiv mm_{I}/\left(m+m_{I}\right)$.

In a standard mean-field approach of studying the BCS-BEC crossover,
the superfluid Hamiltonian can be approximated by 
\begin{equation}
\mathcal{H}_{{\rm sf}}\approx\langle\mathcal{H}_{{\rm sf}}\rangle_{{\rm MF}}+\sum_{\sigma=\uparrow,\downarrow}\sum_{\mathbf{k}}E_{\mathbf{k}}\alpha_{\mathbf{k}\sigma}^{\dagger}\alpha_{\mathbf{k}\sigma},
\end{equation}
where the mean-field ground state energy 
\begin{equation}
\langle\mathcal{H}_{{\rm sf}}\rangle_{{\rm MF}}=-\frac{\Delta^{2}}{g}+\sum_{\mathbf{k}}(\epsilon_{\mathbf{k}}-\mu-E_{\mathbf{k}})=-\frac{m\Delta^{2}}{4\pi\hbar^{2}a}+\sum_{\mathbf{k}}(\epsilon_{\mathbf{k}}-\mu-E_{\mathbf{k}}+\frac{\Delta^{2}}{2\epsilon_{\mathbf{k}}})
\end{equation}
is a constant and will be neglected hereafter. Here, $\Delta=g\sum_{\mathbf{k}}\langle\psi_{-\mathbf{k}\downarrow}\psi_{\mathbf{k}\uparrow}\rangle$
is the mean-field pairing gap. The quasi-particle excitation spectrum
$E_{\mathbf{k}}=\sqrt{\xi_{\mathbf{k}}^{2}+\Delta^{2}}\equiv\sqrt{(\epsilon_{\mathbf{k}}-\mu)^{2}+\Delta^{2}}$
and the corresponding creation (annihilation) operators $\alpha_{\mathbf{k}\sigma}^{\dagger}$
($\alpha_{\mathbf{k}\sigma}$) can be obtained by a Bogoliubov transformation
\begin{equation}
\left(\begin{array}{c}
\alpha_{\mathbf{k}\uparrow}\\
\alpha_{-\mathbf{k}\downarrow}^{\dagger}
\end{array}\right)=\left(\begin{array}{cc}
u_{\mathbf{k}} & -v_{\mathbf{k}}\\
v_{\mathbf{k}} & u_{\mathbf{k}}
\end{array}\right)\left(\begin{array}{c}
\psi_{\mathbf{k}\uparrow}\\
\mathrm{\psi}_{-\mathbf{k}\downarrow}^{\dagger}
\end{array}\right),
\end{equation}
where $u_{\mathbf{k}}^{2}=\left[1+\xi_{\mathbf{k}}/E_{\mathbf{k}}\right]/2$,
$v_{\mathbf{k}}^{2}=1-u_{\mathbf{k}}^{2}$, and $u_{\mathbf{k}}v_{\mathbf{k}}=\Delta/\left(2E_{\mathbf{k}}\right)$
are the quasiparticle wavefunctions. With the quasiparticle operators,
the impurity-fermion interaction can be written as,

\begin{equation}
\begin{aligned}\mathcal{H}_{{\rm int}} & =g_{\uparrow}\sum_{\mathbf{k},\mathbf{k}^{\prime},\mathbf{q}}\left(u_{\mathbf{k}}\alpha_{\mathbf{k}\uparrow}^{\dagger}+v_{\mathbf{k}}\alpha_{-\mathbf{k}\downarrow}\right)c_{\mathbf{q}-\mathbf{k}}^{\dagger}c_{\mathbf{q}-\mathbf{k}^{\prime}}\left(u_{\mathbf{k}^{\prime}}\alpha_{\mathbf{k}^{\prime}\uparrow}+v_{\mathbf{k}^{\prime}}\alpha_{-\mathbf{k}^{\prime}\downarrow}^{\dagger}\right)\end{aligned}
.
\end{equation}

To find the polaron energy at momentum $\mathbf{k}$, i.e., $\mathcal{E}_{P}(\mathbf{k})$,
we adopt an extended Chevy's ansatz \citep{Chevy2006} 
\begin{equation}
|\Psi_{P}(\mathbf{k})\rangle=\left(\phi_{0}\hat{\psi}_{\mathbf{k}}^{\dagger}+\sum_{\mathbf{q},\mathbf{p}}\phi_{\mathbf{q},\mathbf{p}}\hat{\psi}_{\mathbf{k}+\mathbf{q}-\mathbf{p}}^{\dagger}\alpha_{\mathbf{p}\uparrow}^{\dagger}\alpha_{\mathbf{-q}\downarrow}^{\dagger}\right)|0\rangle_{I}|{\rm BCS\rangle_{\uparrow\downarrow}},
\end{equation}
which was suggested by Yi and Cui in an investigation of the $g_{\uparrow}=g_{\downarrow}$
case \citep{Yi2015}. Minimizing $\langle\Psi_{P}(\mathbf{k})|\mathcal{H}|\Psi_{P}(\mathbf{k})\rangle$
with respect to the coefficients $\phi_{0}$, $\phi_{\mathbf{q},\mathbf{p}}$
yields the following equations

\begin{equation}
\left[\mathcal{E}_{P}(\mathbf{k})-\epsilon_{\mathbf{k}}^{(I)}\right]\phi_{0}=g_{\uparrow}\sum_{\mathbf{q}}v_{\mathbf{q}}^{2}\phi_{0}+g_{\uparrow}\sum_{\mathbf{q},\mathbf{p}}v_{\mathbf{q}}u_{\mathbf{p}}\phi_{\mathbf{\mathbf{q}},\mathbf{p}},
\end{equation}
\begin{equation}
\left[\mathcal{E}_{P}(\mathbf{k})-\left(\epsilon_{\mathbf{p}+\mathbf{q}-\mathbf{p}}^{(I)}+E_{\mathbf{p}}+E_{\mathbf{q}}\right)\right]\phi_{\mathbf{q},\mathbf{p}}=g_{\uparrow}v_{\mathbf{q}}u_{\mathbf{p}}\phi_{0}+g_{\uparrow}u_{\mathbf{p}}\sum_{\mathbf{p}^{\prime}}u_{\mathbf{p}^{\prime}}\phi_{\mathbf{q},\mathbf{p}^{\prime}}-g_{\uparrow}v_{\mathbf{q}}\sum_{\mathbf{q}^{\prime}}v_{\mathbf{q}^{\prime}}\phi_{\mathbf{q}^{\prime},\mathbf{p}}+g_{\uparrow}\sum_{\mathbf{q}^{\prime}}v_{\mathbf{q}^{\prime}}^{2}\phi_{\mathbf{q},\mathbf{p}}.
\end{equation}
which can be solve self-consistently by introducing an auxiliary function

\begin{equation}
\chi_{\mathbf{q}}=g_{\uparrow}\left(v_{\mathbf{q}}\phi_{0}+\sum_{\mathbf{p}}u_{\mathbf{p}}\phi_{\mathbf{q},\mathbf{p}}\right).\label{eq:chi_q}
\end{equation}
Recall that renormalization would make $g_{\uparrow}$ and terms including
$g_{\uparrow}\sum_{\mathbf{q'}}v_{\mathbf{q}^{\prime}}...$ vanishingly
small (since $v_{\mathbf{q}'}\sim1/q^{\prime2}$), we can express
$\phi_{0}$ and $\phi_{\mathbf{q},\mathbf{k}}$ as

\begin{equation}
\phi_{0}=\frac{1}{\mathcal{E}_{P}(\mathbf{k})-\epsilon_{\mathbf{k}}^{(I)}}\sum_{\mathbf{q}}v_{\mathbf{q}}\chi_{\mathbf{q}},
\end{equation}
\begin{equation}
\phi_{\mathbf{q},\mathbf{p}}=\frac{1}{\mathcal{E}_{P}(\mathbf{k})-\epsilon_{\mathbf{p}+\mathbf{q}-\mathbf{p}}^{(I)}-E_{\mathbf{p}}-E_{\mathbf{q}}}\chi_{\mathbf{q}}u_{\mathbf{p}}
\end{equation}
and insert them back to Eq. (\ref{eq:chi_q}). Finally, we arrive
at the equation

\begin{equation}
\left(\frac{1}{g_{\uparrow}}-\sum_{\mathbf{p}}\frac{u_{\mathbf{p}}^{2}}{\mathcal{E}_{P}(\mathbf{k})-\epsilon_{\mathbf{p}+\mathbf{q}-\mathbf{p}}^{(I)}-E_{\mathbf{p}}-E_{\mathbf{q}}}\right)\chi_{\mathbf{q}}=\frac{v_{\mathbf{q}}}{\mathcal{E}_{P}(\mathbf{k})-\epsilon_{\mathbf{k}}^{(I)}}\sum_{\mathbf{q}^{\prime}}v_{\mathbf{q^{\prime}}}\chi_{\mathbf{q}^{\prime}},
\end{equation}
which gives Eq. (17) in the main text 
\begin{equation}
\begin{aligned}\mathcal{E}_{P}(\mathbf{k})= & \epsilon_{\mathbf{k}}^{(I)}+\sum_{\mathbf{q}}v_{\mathbf{q}}^{2}\left[\frac{1}{g_{\uparrow}}-\sum_{\mathbf{p}}\frac{u_{\mathbf{p}}^{2}}{\mathcal{E}_{P}(\mathbf{k})-\epsilon_{\mathbf{p}+\mathbf{q}-\mathbf{p}}^{(I)}-E_{\mathbf{p}}-E_{\mathbf{q}}}\right]^{-1},\\
= & \epsilon_{\mathbf{k}}^{(I)}+\sum_{\mathbf{q}}v_{\mathbf{q}}^{2}\left[\frac{m_{r}}{2\pi\hbar^{2}a_{\uparrow}}-\sum_{\mathbf{p}}\frac{2m_{r}}{\hbar^{2}p^{2}}+\sum_{\mathbf{p}}\frac{u_{\mathbf{p}}^{2}}{\epsilon_{\mathbf{p}+\mathbf{q}-\mathbf{p}}^{(I)}+E_{\mathbf{p}}+E_{\mathbf{q}}-\mathcal{E}_{P}(\mathbf{k})}\right]^{-1},
\end{aligned}
\end{equation}
after some manipulation of algebra.
\end{widetext}

\end{document}